\documentclass[onefignum,onetabnum,dvipsnames]{siamart171218}

\usepackage{amsfonts}
\expandafter\def\csname ver@luatexbase.sty\endcsname{}
\usepackage{mathtools, accents}
\usepackage{graphicx}
\usepackage[caption=false, subrefformat=parens]{subfig}
\usepackage{epstopdf}
\usepackage{algorithmic}
\usepackage{multirow}
\usepackage{booktabs}
\usepackage{tabularx}
\usepackage{tikz, pgfplots}
\usetikzlibrary{arrows,calc,positioning,decorations.pathmorphing,decorations.pathreplacing,matrix,fit,backgrounds,external}
\usepgfplotslibrary{groupplots,fillbetween,external} 
\pgfplotsset{width=7cm,compat=newest}
\ifpdf
  \DeclareGraphicsExtensions{.eps,.pdf,.png,.jpg}
\else
  \DeclareGraphicsExtensions{.eps}
\fi

\tikzset{every picture/.style=thick}

\definecolor{RYB1}{RGB}{141, 211, 199}
\definecolor{RYB2}{RGB}{255, 255, 179}
\definecolor{RYB3}{RGB}{190, 186, 218}
\definecolor{RYB4}{RGB}{251, 128, 114}
\definecolor{RYB5}{RGB}{128, 177, 211}
\definecolor{RYB6}{RGB}{253, 180, 98}
\definecolor{seagreen}{RGB}{46, 139, 87}
\definecolor{midnight}{RGB}{25,25,112}

\pgfplotscreateplotcyclelist{mycolorlist}{%
	RYB1!50!black,every mark/.append style={fill=RYB1},mark=*\\%
	RYB2!50!black,every mark/.append style={fill=RYB2},mark=square*\\%
	RYB3!50!black,every mark/.append style={fill=RYB3},mark=otimes*\\%
	RYB4!50!black,mark=star,every mark/.append style=solid\\%
	RYB5!50!black,every mark/.append style={fill=RYB5},mark=diamond*\\%
	RYB6!50!black,every mark/.append style={solid,fill=RYB6},mark=*\\%
}

\pgfplotscreateplotcyclelist{colorscale}{%
	seagreen!50!black,every mark/.append style={fill=seagreen},mark=*,mark options={solid},dotted\\%
	seagreen!50!black,every mark/.append style={fill=seagreen},mark=square*,mark options={solid},dashed\\%
	seagreen!50!black,every mark/.append style={fill=seagreen},mark=otimes*,mark options={solid}\\%
	midnight!50!black,every mark/.append style={fill=midnight},mark=*,mark options={solid},dotted\\%
	midnight!50!black,every mark/.append style={fill=midnight},mark=square*,mark options={solid},dashed\\%
	midnight!50!black,every mark/.append style={fill=midnight},mark=otimes*,mark options={solid}\\%
	red!50!black,every mark/.append style={fill=red},mark=*,mark options={solid},dotted\\%
	red!50!black,every mark/.append style={fill=red},mark=square*,mark options={solid},dashed\\%
	red!50!black,every mark/.append style={fill=red},mark=otimes*,mark options={solid}\\%
}

\pgfplotscreateplotcyclelist{timingcolorlist}{%
	SeaGreen!70!white,every mark/.append style={fill=seagreen!70!white,solid},mark=*\\%
	MidnightBlue!70!white,every mark/.append style={fill=midnight!70!white,solid},mark=square*\\%
	BrickRed!70!white,every mark/.append style={fill=red!70!white,solid},mark=triangle*\\%
}

\pgfplotscreateplotcyclelist{breakdowncolorlist}{%
	fill=SeaGreen!50!white, draw=none\\%
	fill=MidnightBlue!50!white, draw=none\\%
	fill=BrickRed!50!white, draw=none\\%
	fill=BurntOrange!50!white, draw=none\\%
	fill=Aquamarine!50!white, draw=none\\%
}

\pgfplotsset{
	every axis/.append style={
		thick,
		line width=1pt
	},
	every axis legend/.append style={
		line width=1pt
	},
	two plot/.style={
		width=0.95*\textwidth,
		height=0.35*\textheight,
		ymajorgrids,
		enlargelimits=false,
		legend columns=3,		
		legend cell align=left,
		legend style={font=\small,draw=none},
	},
	timing plot/.style={
		two plot,
		cycle list name=timingcolorlist,
		ylabel={updates per second (Hz)}
	},
	deform timing plot/.style={
		timing plot,
        xlabel={cumulative constrained node count},
		legend entries={AMPS, $D$ precond.~aug., Jacobi precond.~CG}
	},
	cutting timing plot/.style={
		timing plot,
		xlabel={cumulative cut node count},
	},
    cutting speedup plot/.style={
		timing plot,
		xlabel={cumulative cut node count},
		legend entries={AMPS (single-core), AMPS (32-core)}
	},
	scale plot/.style={
		timing plot,
		xlabel={number of mesh nodes},
		legend entries={AMPS, $D$ precond.~aug., Jacobi precond.~CG}
	},
	line and fill/.style={
		legend image code/.code={%
			\fill [##1,opacity=0.1,draw=none]
				(0mm,-1ex) rectangle (6mm,1ex);
			\draw [##1] (0mm,0mm) -- plot (3mm,0mm) -- (6mm,0mm);
		}
	},
}
\newcommand{\errorband}[6][]{
	\addplot[name path={#2_#4},draw=none,forget plot]
		table [x={#3},y={#4}] {#2};
	\addplot[name path={#2_#5},draw=none,forget plot]
		table [x={#3},y={#5}] {#2};
	\addplot[#1,opacity=0.1,draw=none,forget plot]
		fill between[of={#2_#4} and {#2_#5}];
	\addplot[#1,line width=1pt,fill=none,forget plot]
		table [x={#3},y={#6}] {#2};
	\addlegendimage{line and fill,#1};
}

\newcommand{\TheTitle}{AMPS:  Real-time Mesh Cutting with Augmented Matrices for Surgical Simulations} 
\newcommand{\TheAuthors}{Y.-H. Yeung, A. Pothen, and J. Crouch}

\headers{Real-time Mesh Cutting with Augmented Matrices}{\TheAuthors}

\title{{\TheTitle}\thanks{This work was supported in part by 
NSF grant  CCF-1637534; the 
U.S. Department of Energy through grant DE-FG02-13ER26135; and  the Exascale Computing Project (17-SC-20-SC), a collaborative effort of the DOE Office of Science and the NNSA.  
}}

\author{
  Yu-Hong Yeung\thanks{Department of Computer Science, Purdue University, West Lafayette, IN
		(\email{yyeung@purdue.edu}, \email{apothen@purdue.edu}).}
  \and
	Alex Pothen\footnotemark[2]
	\and
	Jessica Crouch\thanks{Department of Computer Science, Old Dominion University, Norfolk, VA
	(\email{jrcrouch@cs.odu.edu}).}
}

\usepackage{amsopn}

\newcommand\closure[2]{\mathop{\mathrm{closure}_{#1}\left(#2\right)}}

\newcommand\tril[1]{\operatorname{\mathrm{tril}}\left(#1\right)}

\ifpdf
\hypersetup{
  pdftitle={AMPS: A Real-time Mesh Cutting Algorithm for Surgical Simulations},
  pdfauthor={Y.-H. Yeung, A. Pothen, and J. Crouch}
}
\fi

\begin{document}

\maketitle

\begin{abstract}
We present the AMPS algorithm,  a finite element solution method that combines  principal submatrix updates and Schur complement techniques,  well-suited for interactive simulations of deformation and cutting of finite element meshes.  Our approach features real-time solutions to the updated stiffness matrix systems to account for interactive
changes in mesh connectivity and boundary conditions.  Updates are
accomplished by an augmented matrix formulation of the stiffness equations to maintain its consistency with changes to the underlying model without refactorization at each timestep.  As changes accumulate over multiple simulation timesteps, the augmented solution algorithm  enables tens or hundreds of updates per second. 
Acceleration schemes that exploit sparsity, memoization and parallelization lead to the  updates being computed in real-time.  The complexity analysis and experimental results for this
method demonstrate that it scales linearly  with the problem size.  Results for cutting and deformation of 3D elastic models are reported for meshes with node counts up to 50,000, and involve models of astigmatism surgery and the brain. 
\end{abstract}



\section{Introduction}
We present an algorithm to support real-time deformation and cutting of solid
finite element models by quickly solving the resulting time-varying equations.
Topological mesh modifications and boundary condition changes are the basic
operations of many simulation scenarios, particularly surgical simulations.
Real-time finite element solution methods for mesh cutting is a
computational challenge, first because graphic and haptic rendering demand
accurate solutions at real-time update rates, and second because connectivity
changes due to cutting and remeshing modifies the underlying matrix equations.
Such modifications invalidate previous factorizations or inverse computations
for the stiffness matrix, requiring either computationally expensive update
procedures or solution via an iterative method.

Interactive simulations often involve unpredictable cutting paths to allow
flexibility to the user inputs. This feature requires that the internal
deformation of a solid model be computed and tracked so that accurate cut
surfaces are exposed as cuts progress into a model's potentially
heterogeneous interior. While the 3D mesh is changed due to cutting,
pushing and pulling forces are being applied, and new Dirichlet boundary
conditions are being imposed by different fixation scenarios, a real-time
solution method to compute the displacement of all nodes under these
changes is essential to make the simulations practical.

Observing  that the aforementioned changes to the meshes result
in a principal submatrix update and a change in dimensions to the underlying
equations, we propose a new solution approach to reflect both the update and
the dimension change in a modified augmented matrix  formulation. This approach is similar to other augmented matrix methods in that the matrix is represented in a block matrix form in which the (1,1) block is the fixed original matrix and the other blocks are either zero or vary according to the changes. 
The Schur complement operation is then applied to decouple the augmentation from the remaining part of the system, and the Schur complement system is solved in two phases. 
Our current solution  combines a one-time sparse
matrix-factorization  for the (1,1) block with an explicit
computation of a principal submatrix of the inverse of the original matrix
and a direct solution of the Schur complement system. 
Sparsity in the matrix, solution vector, and the right-hand-side vector are carefully exploited throughout the computations and intermediate results are stored for subsequent changes in later cutting steps. The time complexity of the algorithm shows that performance scales well with model size
and various cutting lengths,  while supporting arbitrary cutting of any valid finite element mesh.

Different algorithms for mesh generation \cite{crouch} \cite{goksel2011}
\cite{lederman} \cite{mohamed}, collision detection \cite{spillmann}
\cite{teschner} \cite{zhang}, and mesh refinement \cite{forest}
\cite{mor} \cite{steinemann} can be paired with our solution algorithm to 
produce a complete simulation platform. Thus the scope of this paper does not include algorithms for simulation tasks other than solving the finite element system of equations.  A feature of the solution algorithm presented is its flexibility to work with structured and unstructured meshes as well as a number of different methods for adapting mesh geometry to respect a cut surface.

The three main contributions of this work are:
\begin{itemize}
	\item An augmented matrix formulation of the stiffness system of equations from a finite element model, 
	specific for principal submatrix updates and dimension changes resulting
	from both continuous unpredictable cutting and imposition of new boundary conditions. This formulation keeps the original stiffness matrix as a submatrix to eliminate the necessity of re-factorization whenever a change occurs.
	\item A direct solution approach that provides fast and accurate
	solutions to both the updated portion and unchanged portion, when the
	percentage of mesh elements affected by topological changes is small.
	\item Acceleration of the solution method by exploiting sparsity,
    memoization and parallelization. We analyze the time complexity of the
    accelerated solution method using concepts from graph theory.
\end{itemize}

This paper is organized as follows. Section~\ref{sec:previous_work} reviews
previous work on the real-time solution of physics-based models and finite
element equations. Section~\ref{sec:methods} presents our new augmented
method with principal submatrix update for assembling a finite element system of equations and accounting for changes in mesh connectivity and boundary conditions via updates to stiffness matrix factors. Section~\ref{sec:results}
presents speed and accuracy results from finite element deformation and 
cutting experiments with models of various size. Finally, 
Section~\ref{sec:conclusion} discusses conclusions and directions for future 
work.

\section{Previous Work}
\label{sec:previous_work}
The augmented matrix algorithm presented in this paper is related
to those presented by us and our colleagues in \cite{yeung16} and
\cite{yeung17}. In the first paper, we formed an augmented system to
replace columns in the original matrix, and solved the Schur complement
system using GMRES implicitly and the rest of the system directly using
precomputed $LDL^\top$ factors of the original matrix. Symmetry of the
system was destroyed during the update, and thus two closures needed to be computed to exploit the sparsities in both the matrix and the right-hand-side vector.
The convergence of the iterative solver depended on the condition of the Schur complement of the system, and a preconditioner was sometimes needed for faster convergence. However, the absence of the explicit Schur complement made finding a fast and efficient preconditioner difficult.

To overcome these shortcomings, we follow an approach similar to that
presented in the second paper. By observing that the only change to the
original matrix is within a principal submatrix, with our co-authors we
showed that symmetry could be preserved during the update. We presented
two approaches to solve the Schur complement system, an iterative method
and a direct method. However, the  contingency analysis application for power grids considered there  retained the size of the system for any contingency scenario.
Thus  the augmented system considered there applied  to  applications that
do not change the matrix dimension. This is not the case with surgical
simulations, in which new vertices are added to the mesh along the cutting
surface. The additional vertices increase the overall dimension of the
modified system. An extension is, therefore, presented in this paper to
generalize the augmented matrix approach to systems where their dimensions
change. We also improve the computation of
the principal submatrix of the matrix inverse to further accelerate the
solution.

In \cite{yeung17}, CHOLMOD~\cite{davis}, an algorithm to update or downdate the
Cholesky factor of the matrix with low-rank matrices, was compared to
our augmented matrix formulation. It was shown that our approach
outperformed CHOLMOD for the power contigency application. However,
SuiteSparse, the software package that includes CHOLMOD, does not
provide functionality to increase the dimension of the modified system.
We, therefore, do not provide the comparison between our method and
CHOLMOD for the surgical simulation application in this paper.

Other related papers were surveyed in the two aforementioned papers and
hence we do not repeat them here.


\section{Methods}
\label{sec:methods}
In this paper we denote by $n$ the order of the original matrix, $m_t$ the number of
its rows and columns replaced at time $t$, and $k_t$ the change in dimension of the
modified matrix at time $t$. Hence the modified matrix has order $(n+k_t)$. The original
stiffness system is $K a = f$, where the right-hand-side vector $f$ is called the force
vector. In the context of the finite element model used in the surgical simulation,
$m_t$ corresponds to the degrees of freedoms (DOFs) of the modified vertices and their
neighbors, and $k_t$ corresponds to the DOFs of the newly added vertices with respect to 
the original system. In general, $m_t \gg k_t$.

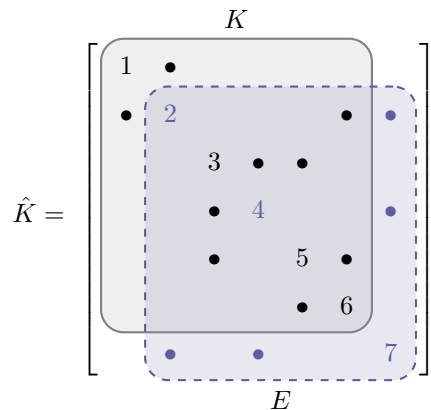
\begin{figure}[ht!]
	\centering
	\begin{tikzpicture}
		\matrix[matrix of math nodes,
        	column sep=0.5em,
            row sep=0.5em,
			left delimiter={[},
            right delimiter={]}
            ] (K) {
			1 & \bullet & & & & & \\
			\bullet & \textcolor{midnight!70}{2} & & & & \bullet & \textcolor{midnight!70}{\bullet} \\
			& & 3 & \bullet & \bullet & & \\
			& & \bullet & \textcolor{midnight!70}{4} & & & \textcolor{midnight!70}{\bullet} \\
			& & \bullet & & 5 & \bullet & \\
			& & & & \bullet & 6 & \\
			& \textcolor{midnight!70}{\bullet} & & \textcolor{midnight!70}{\bullet} & & & \textcolor{midnight!70}{7}\\	
		};
		\node [left, xshift=-1em] at (K.west) {$\hat{K} =$};
		\begin{pgfonlayer}{background}
			\node[draw=black!50,
            	thick,
				fill=black!30,
                fill opacity=0.2,
            	rounded corners=2ex,
				label=above:$K$,
				fit=(K-1-1.north west)(K-6-6.south east)] {} ;
			\node[draw=midnight!70,
				thick, dashed,
                fill=midnight!50,
                fill opacity=0.2,
                rounded corners=2ex,
				label=below:$E$,
				fit=(K-2-2.north west)(K-7-7.south east)] {} ;
		\end{pgfonlayer}
	\end{tikzpicture}
	\caption{Example of a modified $(7 \times 7)$-matrix $\hat{K}$
    formed by a principal submatrix update $E$ of size $3 \times 3$
    enclosed in blue dashed box to the original $(6 \times 6)$-matrix $K$ 
    enclosed in grey solid box with dimension change.}
    \label{fig:psud_eg}
\end{figure}

By considering the difference between the original $n \times n$ stiffness matrix $K$ and
the modified $(n + k_t) \times (n + k_t)$ stiffness matrix $\hat{K}$ after cutting at time
$t$ as illustrated in Figure~\ref{fig:psud_eg}, we observe that $\hat{K}$ can be expressed
as the result of a principal submatrix update to $K$ augmented by an identity matrix of
size $k_t$:
\begin{equation}
	\hat{K} = \underbrace{\begin{bmatrix}
		K & \\ & I_{k_t}
	\end{bmatrix}}_{\displaystyle\bar{K}} -
	\underbrace{\begin{bmatrix}
		H & \\ & I_{k_t}
	\end{bmatrix}}_{\displaystyle\bar{H}}
	\underbrace{\left(E + \begin{bmatrix}
		0_{m_t} & \\ & I_{k_t}
	\end{bmatrix}\right)}_{\displaystyle\bar{E}}
	\underbrace{\begin{bmatrix}
		H^\top & \\ & I_{k_t}
	\end{bmatrix}}_{\displaystyle\bar{H}^\top},
	\label{eq:update}
\end{equation}
where $H$ comprises the $m_t$ columns of the identity matrix of size $n$ whose indices
correspond to the rows and columns of $K$ to be updated; and $E$ is an
$(m_t + k_t) \times (m_t + k_t)$ principal submatrix update to $\bar{K}$. Here $H$ has
dimension $n \times m_t$, $\bar{H}$ has dimension $(n + k_t) \times (m_t + k_t)$; and
$\bar{E}$ has dimension $(m_t + k_t) \times (m_t + k_t)$, the same as that of $E$. Note
that $\bar{H}^\top \bar{H} = I_{(m_t + k_t)}$.

If we express $\hat{a}$ as the sum of two independent terms:
\begin{equation}
    \hat{a} = \bar{a}_1 + \bar{H}\bar{a}_2
    \label{eq:ahat}
\end{equation}
such that
\begin{equation}
    \bar{H}^\top\bar{a}_1 = 0,
\end{equation}
which implies $\bar{H}^\top\hat{a} = \bar{a}_2$, and let
\begin{equation}
    \bar{a}_3 = \bar{H}^\top
        \left(\bar{f} - \hat{f}\right) -
        \bar{E}\bar{a}_2,
\end{equation}
then with some arithmetic operations, we can show that a principal submatrix update
in the form of Equation~\ref{eq:update} can be solved using an augmented matrix 
formulation
\begin{equation}
	\begin{bmatrix}
		\bar{K} & \bar{K}\bar{H} & \bar{H}\\
		\bar{H}^\top\bar{K} & \bar{H}^\top \bar{K} \bar{H} - \bar{E} & 0\\
		\bar{H}^\top & 0 & 0
	\end{bmatrix} \begin{bmatrix}
		\bar{a}_1 \\ \bar{a}_2 \\ \bar{a}_3
	\end{bmatrix} = \begin{bmatrix}
		\bar{f} \\ \bar{H}^\top \hat{f} \\ 0
	\end{bmatrix},\label{eq:aug}
\end{equation}
where $\bar{f}$ is the $(n + k_t)$-vector obtained by padding $k_t$ zeros at the end of
the force vector $f$, and $\hat{f}$ is the force vector applied to the modified mesh.
With $\bar{K}$ as the pivot, Equation~\ref{eq:aug} can be reduced to a smaller system
involving the symmetric matrix $S_1$, the Schur complement of $\bar{K}$, and after
multiplication with $-1$ we obtain:
\begin{equation}
	\underbrace{\begin{bmatrix}
		\bar{E} & I\\
		I & \bar{H}^\top\bar{K}^{-1}\bar{H}
	\end{bmatrix}}_{\displaystyle S_1} \begin{bmatrix}
		\bar{a}_2 \\ \bar{a}_3
	\end{bmatrix} = \begin{bmatrix}
		\bar{H}^\top \left(\bar{f} - \hat{f}\right) \\
		\bar{H}^\top\bar{K}^{-1}\bar{f}
	\end{bmatrix},\label{eq:s1}
\end{equation}
in which
\begin{equation}
	\bar{K}^{-1} = \begin{bmatrix}
		K^{-1} & \\
		& I_k
	\end{bmatrix}.\label{eq:Kinv}
\end{equation}
Equation~\ref{eq:s1} can be further reduced with another Schur complement using the
$(1,2)$-block of $S_1$ as the block pivot:
\begin{align}
	\underbrace{
	    \left(I-\bar{H}^\top\bar{K}^{-1}\bar{H}\bar{E}\right)
	}_{\displaystyle S_2} \bar{a}_2
		&= \bar{H}^\top \bar{K}^{-1}\bar{f}
		-\bar{H}^\top \bar{K}^{-1}\bar{H} \bar{H}^\top
		\left(\bar{f}-\hat{f}\right)\nonumber\\
		&= \bar{H}^\top \bar{K}^{-1}\hat{f}.
		\label{eq:s2}
\end{align}
Note that the matrix $S_2$ is not symmetric. If $\hat{f}$ only differs from $f$ at the
newly added vertices, i.e.
\begin{equation}
	\hat{f} - \bar{f} = \begin{bmatrix}
		0_n \\ \accentset{\circ}{f}
	\end{bmatrix},\label{eq:fcond}
\end{equation}
then the right-hand-side vector of Equation~\ref{eq:s2} can be simplified to
\begin{equation}
	\bar{H}^\top\bar{K}^{-1}\hat{f} = \begin{bmatrix}
		H^\top a\\\accentset{\circ}{f}
	\end{bmatrix},\label{eq:s2rhs}
\end{equation}
where $a$ is the solution to the original system $Ka=f$ and $\accentset{\circ}{f}$ is
the force applied to the $k_t$ newly added vertices.

After solving Equation~\ref{eq:s2} for $\bar{a}_2$ using a direct solver, we can solve
for $\hat{a}$ in the modified system $\hat{K}\hat{a}=\hat{f}$ directly using the
following observation. Premultiplying the first row block of Equation~\ref{eq:aug} by
$\bar{K}^{-1}$ and rearranging the terms yields
\begin{equation}
	\bar{a}_1 = \bar{K}^{-1}\bar{f} - \bar{H}\bar{a}_2 - \bar{K}^{-1}\bar{H}\bar{a}_3.
	\label{eq:augb1}
\end{equation}
In addition, rearranging the terms in first row block of Equation~\ref{eq:s1} yeilds
\begin{equation}
	\bar{a}_3 = \bar{H}^\top \left(\bar{f} - \hat{f}\right) - \bar{E}\bar{a}_2.
	\label{eq:s2b1}
\end{equation}
Substituting Equation~\ref{eq:s2b1} into Equation~\ref{eq:augb1} yields
\begin{equation}
	\bar{a}_1 = \bar{K}^{-1}\left[\bar{f}
			-\bar{H}\bar{H}^\top\left(\bar{f}
        - \hat{f}\right)\right]
		+\left(\bar{K}^{-1}\bar{H}\bar{E}
        - \bar{H}\right)\bar{a}_2.\label{eq:a1}
\end{equation}
Again, if $\hat{f}$ satisfies the condition of Equation~\ref{eq:fcond},
Equation~\ref{eq:a1} can be simplified to
\begin{equation}
	\bar{a}_1 = \begin{bmatrix}
		a \\ \accentset{\circ}{f}
	\end{bmatrix}
		+\left(\bar{K}^{-1}\bar{H}\bar{E}-\bar{H}\right)\bar{a}_2.\label{eq:a1sim}
\end{equation}
Substituting Equation~\ref{eq:a1sim} into Equation~\ref{eq:ahat} yields
\begin{equation}
	\hat{a} = \begin{bmatrix}
		a \\ \accentset{\circ}{f}
	\end{bmatrix} +
	\bar{K}^{-1}\bar{H}\bar{E}\bar{a}_2,
	\label{eq:sol_a2}
\end{equation}
thus completing the solution.

An alternative Schur complement formulation is possible. One can use the $(2,1)$-block
in Equation~\ref{eq:s1} as the block pivot for the Schur complement and get
\begin{align}
	\left(\bar{E}\bar{H}^\top\bar{K}^{-1}\bar{H}-I\right)\bar{a}_3
		&= \bar{H}^\top \left(\hat{f} - \bar{f}\right)
		+ \bar{E}\bar{H}^\top\bar{K}^{-1}\bar{f}.\nonumber\\
		&= \begin{bmatrix}
			0 \\ \accentset{\circ}{f}
		\end{bmatrix} + \begin{bmatrix}
			E_{11} \\ E_{12}^\top
		\end{bmatrix}H^\top a,\label{eq:s3}
\end{align}
assuming that the condition in Equation~\ref{eq:fcond} is satisfied. Again the
coefficient matrix is not symmetric. After solving for $\bar{a}_3$ using
Equation~\ref{eq:s3}, the solution $\hat{a}$ can be obtained as follows:
\begin{equation}
	\hat{a} = \begin{bmatrix}
		a \\ \accentset{\circ}{f}
	\end{bmatrix} - \bar{K}^{-1}\bar{H}\bar{a}_3.
\end{equation}

\subsection{Improving numerical accuracy}
We can improve the numerical accuracy of the solutions by substituting $\bar{a}_2$
into $\hat{a}$ directly as follows. From the third row block of Equation~\ref{eq:aug},
we have
\begin{equation}
	\bar{H}^\top\bar{a}_1 = 0.
\end{equation}
Premultiplying Equation~\ref{eq:ahat} by $\bar{H}^\top$ yields
\begin{equation}
	\bar{H}^\top\hat{a} = \bar{H}^\top\bar{a}_1 + \bar{H}^\top\bar{H}\bar{a}_2
	= \bar{a}_2.
\end{equation}
Note that the components of $\hat{a}$ picked out by $\bar{H}^\top$ correspond to
$\bar{a}_2$, which are arithmetically identical to the same components computed using
Equation~\ref{eq:sol_a2} but with higher accuracy. If we denote $\mathbb{H}$ as the set
of indices for which the rows and columns of $A$ are updated including the newly added
ones, combining the two equations, we have
\begin{equation}
	\hat{a}[i] = \begin{cases}
		\left(\bar{H}\bar{a}_2\right)[i] &\text{for }i\in\mathbb{H},\\
		\left(\begin{bmatrix}
		a \\ \accentset{\circ}{f}
	\end{bmatrix} + \bar{K}^{-1}\bar{H}\bar{E}\bar{a}_2\right)[i]
    	& \text{for }i\notin\mathbb{H}.
	\end{cases}.\label{eq:hatai}
\end{equation}
Skipping the computations of those components in $\hat{a}$ that are in $\mathbb{H}$
also improves the performance of the algorithm.

\subsection{Computing the Schur Complement Matrix}
Our augmented algorithm involves solving Equations~\ref{eq:s2} and~\ref{eq:hatai}.
Unlike \cite{yeung16} both equations are solved using a direct solver. The Schur
complement matrix $S_2$ in Equation~\ref{eq:s2} can be expressed in block matrix form
using Equations~\ref{eq:update}, \ref{eq:Kinv} and~\ref{eq:s2rhs} to obtain
\begin{equation}
	\left(\begin{bmatrix}
		I_m \\ & 0_k
	\end{bmatrix} - \begin{bmatrix}
		H^\top K^{-1} H \\ & I_k
	\end{bmatrix} E\right)\bar{a}_2 = \begin{bmatrix}
		H^\top a \\ \accentset{\circ}{f}
	\end{bmatrix}.\label{eq:block}
\end{equation}
Solving Equation~\ref{eq:block} involves computing the principal submatrix 
of the inverse $H^\top K^{-1} H$. Assuming that $K = LDL^\top$ is a
factorization of $K$, we have
\begin{equation}
	H^\top K^{-1} H = H^\top L^{-\top}D^{-1}L^{-1}H.
\end{equation}
If we denote $V \equiv L^{-1}H$, then $H^\top K^{-1} H = V^\top D^{-1} V$, which can be
computed by first solving for $V$ using forward substitution, then scaling $V$ to obtain
$U \equiv D^{-1} V$ and finally premultiplying $U$ by $V^\top$. The computation of the
rest of the matrix in Equation~\ref{eq:block} is straight forward.

\subsection{Memoization}
For an efficient computation of the principal submatrix of the inverse
$H_t^\top K^{-1} H_t$ at time $t$, we observe that since the vertices removed during the
cutting are accumulating and $H$ is the submatrix of the identity corresponding to the
replaced rows and columns in $K$, the matrix $H_{t-1}$ at the previous time $t-1$ is a
submatrix of the first $m_{t-1}$ columns of matrix $H_t$ at time $t$, i.e.,
\begin{equation}
	H_t = \left[\begin{array}{c|c}
		H_{t-1} & H_{\Delta t}
	\end{array}\right],
\end{equation}
where $H_{\Delta t}$ is the $(m_t - m_{t-1})$ columns of the identity matrix
corresponding to the newly removed columns at timestep $t$. Consequently, 
the matrix $V_{t-1}$ is also the first $m_{t-1}$ columns of $V_t$ since each
column of $V_t$ is independently solved, i.e.,
\begin{equation}
	V_t = \left[\begin{array}{c|c}
		V_{t-1} & V_{\Delta t}
	\end{array}\right],
\end{equation}
where $V_{\Delta t} = L^{-1} H_{\Delta t}$, which are the only columns of 
$V_t$ that need to be computed. Furthermore, the top-left
$(m_{t-1} \times m_{t-1})$ submatrix of $H_t^\top K^{-1} H_t$ is identical 
to $H_{t-1}^\top K^{-1} H_{t-1}$ because
\begin{align}
    H_t^\top K^{-1} H_t &= V^\top_t D^{-1} V_t =
    \left[\renewcommand{\arraystretch}{1.5}\begin{array}{c}
		V^\top_{t-1} \\\hline V_{\Delta t}^\top
	\end{array}\right] D^{-1} \left[\begin{array}{c|c}
		V_{t-1} & V^\top_t
	\end{array}\right]\nonumber\\
	&= \left[\renewcommand{\arraystretch}{1.5}\begin{array}{c|c}
		V^\top_{t-1}D^{-1}V_{t-1} & V^\top_{t-1}D^{-1}V_{\Delta t}\\\hline
		V_{\Delta t}^\top D^{-1}V_{t-1} & V_{\Delta t}^\top D^{-1} V_{\Delta t}
	\end{array}\right]\nonumber\\
	&= \left[\renewcommand{\arraystretch}{1.5}\begin{array}{c|c}
		H_{t-1}^\top K^{-1} H_{t-1} & V^\top_{t-1}D^{-1}V_{\Delta t}\\\hline
		V_{\Delta t}^\top D^{-1}V_{t-1} & V_{\Delta t}^\top D^{-1} V_{\Delta t}
	\end{array}\right].\label{eq:KtH}
\end{align}
Furthermore, it can be observed from Equation~\ref{eq:KtH} that 
$H_t^\top K^{-1} H_t$ is also symmetric and only the lower or upper 
triangular part needs to be computed and stored, and subsequent updates can 
be done sequentially by trapezoidal augmentations to
$\tril{H_{t-1}^\top K^{-1} H_{t-1}}$:
\begin{equation}
    \begin{tikzpicture}[baseline={(x.base)},scale=1.5]
		\draw (0,0) -- (0,3) -- (3,0) -- cycle;
		\draw[dashed] (0,0.8) -- (2.2,0.8);
		\draw [decorate,decoration={brace,amplitude=4pt},xshift=2pt]
			(3,3) -- (3,0.8) node [midway,right,xshift=6pt]{$m_{t-1}$};
		\draw [decorate,decoration={brace,amplitude=4pt},xshift=2pt]
			(3,0.8) -- (3,0) node [midway,right,xshift=6pt]{$m_t - m_{t-1}$};
		\node[anchor=east] at (-0.05,1.2) {$\tril{H_t^\top K^{-1} H_t} =$};
		\node[anchor=west, fill=white] (x) at (0.05,1.2) {$\tril{H_{t-1}^\top K^{-1} H_{t-1}}$};
		\node[anchor=west] at (0.05,0.4) {$\tril{H_{\Delta t}^\top K^{-1} H_t}$};
	\end{tikzpicture},
\end{equation}
where $\tril{\bullet}$ is the lower triangular part of the matrix and the 
augmentation part, $\tril{H_{\Delta t}^\top K^{-1} H_t}$, can be computed as
\begin{multline}
	\tril{H_{\Delta t}^\top K^{-1} H_t}[i,j]
		= \left(V_{\Delta t}[i,*]\right)^\top D^{-1} \left(V_t[j,*]\right)\\
		\text{for }i \in [m_{t-1}+1, m_t]; j \in [1,i].\label{eq:KtHp}
\end{multline}
It is obvious that Equation~\ref{eq:KtHp} can be computed in  parallel for all $i$'s and
$j$'s since they are independent of each other.

\subsection{Dimension Shrinking}
In the case of the imposition of Dirichlet boundary conditions, the 
dimension of the system is shrunk instead of expanded, unlike the case of 
cutting. The authors in \cite{yeung16} have shown that an augmented matrix 
system similar to Equation~\ref{eq:aug} is equivalent to the modified system 
of equations:
\begin{equation}
	\begin{bmatrix}
		K & H\\
		H^\top & 0
	\end{bmatrix} \begin{bmatrix}
		a_1 \\ a_2
	\end{bmatrix} = \begin{bmatrix}
		\hat{f} \\ 0
	\end{bmatrix},\label{eq:deform}
\end{equation}
where $a_2 = -H^\top f$ is the newly unknown force and $\hat{f}$ is given by
\begin{equation}
	\hat{f}[i] = \begin{cases}
		f[i] - 	\displaystyle\sum_{j\in\mathbb{H}} K[i,j]a[j] & j\notin\mathbb{H},\\
		- \displaystyle\sum_{j\in\mathbb{H}} K[i,j]a[j] & j\in\mathbb{H}.
	\end{cases}
\end{equation}
Similar to Equation~\ref{eq:aug}, we can reduce Equation~\ref{eq:deform}
to a smaller system using $K$ as the pivot:
\begin{equation}
	H^\top K^{-1} H a_2 = H^\top K^{-1} \hat{f}.
\end{equation}
Note that the matrix on the left-hand side is the principal submatrix of the
inverse $K^{-1}(H)$, which can be efficiently computed as described in 
previous subsections. The right-hand side can be computed using
$V \equiv L^{-1} H$ as
\begin{equation}
	H^\top K^{-1} \hat{f} = V^\top \underbrace{D^{-1} L^{-1} \hat{f}}_{\displaystyle g}.
    \label{eq:g}
\end{equation}
After computing $a_2$, $a_1$ can be computed using the first row block
of Equation~\ref{eq:deform} as
\begin{align}
	a_1 &= K^{-1} \left(\hat{f} - H^\top a_2\right)\nonumber\\
		&= L^{-\top} D^{-1} L^{-1} \left(\hat{f} - H^\top a_2\right)\nonumber\\
		&= L^{-\top} \left(g - D^{-1} V a_2\right),
\end{align}
in which $g$ is already computed in Equation~\ref{eq:g} and can be reused.

\subsection{Complexity Analysis}
The time complexity of principal submatrix updates using the symmetric augmented
formulation can be summarized in Table~\ref{tab:summary}. Both per cut and total
update times are provided. The one-time factorization costs assume meshes with good
separators for both $2$D  and $3D$-meshes, of size $O(n^{1/2})$ and $O(n^{2/3})$,
respectively. In the table, variables with subscript $t$ are the values at time $t$,
those with subscript $\Delta t$ are the newly added values at time $t$, whereas those
without any subscript are their maxima over all $t$. Recall that $n$ is the size of the
original matrix $K$, $m$ is the size of the principal submatrix update $H$,
$k$ is the dimension change. In addition, $\mathbb{H}$ is the set of indices
of the nonzero rows of $H$, $|L|$ is the number of nonzeros in $L$, $c$ is
the total number of cuts, and $v = \max_j |V_{*,j}|$ is the maximum number
of nonzeros in any column of $V$, which is equivalent to the maximum
closure size of any vertex in the graph of $G(L)$. For a detailed discussion 
on the concepts of closure and the relations between sparse matrix 
computations and its corresponding graph, we refer the readers to \cite{yeung16}. 
The authors also discussed the theorems that are used to prove the upper 
bounds of the complexity of the AMPS algorithms.
\begin{table*}[!ht]
	\begin{tabularx}{\textwidth}
		{l>{\raggedright}p{3cm}>{\centering}p{4.5cm}|>{\centering\arraybackslash}X}
		\toprule
		& Computation & \multicolumn{2}{c}{Complexity} \\\midrule
		    \multicolumn{4}{l}{Amortized initialization:}\\
		1 & Compute LDL$^\top$ factorization of $K$ &
			\multicolumn{2}{c}{$O(n^2)$ for 3D meshes; $O(n^{3/2})$ for 2D meshes}\\
		2 & Compute $a = K^{-1} f$ & \multicolumn{2}{c}{$O(|L|)$}\\\midrule
		\multicolumn{2}{l}{ Real-time update steps:} & per step & total \\
		1 & Solve for $V_{\Delta t}$ &
		    $O\left(\sum_{h\in\mathbb{H}_{\Delta t}}{\closure{L}{h}}\right)$ &
		    $O\left(\sum_{h\in\mathbb{H}}{\closure{L}{h}}\right)$\\
		2 & Compute $\tril{H_{\Delta t}^\top K^{-1} H_t}$ &
		    $O\left((m_{t-1} + 1 + m_t)m_{\Delta t}\cdot v_t\right)$ & $O(m^2v)$\\
		3 & Form $S_2$ & $O\left(m_t^2(m_t + k_t)+m_t\right)$ & $O\left(m^2(m+k)\right)$\\
		4 & Form R.H.S. of Equation~\ref{eq:s2rhs} & $O(m_t)$ & $O(m)$\\
		5 & Solve for $\bar{a}_2$ in Equation~\ref{eq:s2} &
		    $O\left((m_t+k_t)^3\right)$ & $O\left(c\cdot(m+k)^3\right)$\\
		6 & Solve for $\hat{a}$ in Equation~\ref{eq:hatai} &
		    $O(m_t\cdot v_t + |L|)$ & $O\left(c\cdot(mv + |L|)\right)$\\
		\bottomrule
	\end{tabularx}
	\caption{Summary of time complexity}\label{tab:summary}
\end{table*}

The overall time complexity of the algorithm is dominated by either Step~2 (computing
$\tril{H_{\Delta t}^\top K^{-1} H_t}$) or Step~6 (solving for $\hat{a}$). The update
steps in the AMPS algorithm have an overall time complexity of 
\begin{equation}
	O\left(m^2v + c\cdot|L|\right).
\end{equation}

\subsection{Parallelization}
We can observe that Steps~1--4 in the update steps in Table~\ref{tab:summary} are easily
parallelizable from the facts that in Step~1 each columns of $V_{\Delta t}$ are
independently solved, both Steps~2 and 3 involve matrix-matrix multiplications, and in 
Step~4 the R.H.S. of Equation~\ref{eq:s2rhs} is formed by mapping. The parallelization
of Step 5 and 6 is non-trivial, which is out of the scope of this paper. The parallel
time complexity of the update steps in the algorithm for $p$ processors is
\begin{equation}
	O\left(\frac{m^2v}{p}+c\cdot|L|\right).
\end{equation}

\subsection{Relation to previous augmented formulation}
The authors in \cite{yeung16} presented a hybrid asymmetric augmented algorithm to
perform a surgical simulation using finite element models as we do. In their
formulation, the system is augmented in an unsymmetric manner:
\begin{equation}
	\begin{bmatrix}
		\bar{K} & J\\
		\bar{H}^\top & 0
	\end{bmatrix} \begin{bmatrix}
		\bar{a}_1 \\ \bar{a}_2
	\end{bmatrix} = \begin{bmatrix}
		\hat{f} \\ 0
	\end{bmatrix},\label{eq:asymm}	
\end{equation}
where $J$ consists of  the $(m + k)$ columns of $\hat{K}$ to replace the corresponding
columns of $\bar{K}$. Note that we use $\bar{H}^\top$ here for matrices with more columns
than rows instead. They then split Equation~\ref{eq:asymm} into two parts to solve for
$\bar{a}_1$ and $\bar{a}_2$ respectively:
\begin{subequations}
	\begin{align}
        \bar{H}^\top\bar{K}^{-1}J\bar{a}_2 &=
            \bar{H}^\top \bar{K}^{-1}\hat{f}\quad\text{and}\label{eq:prev1}\\
        \bar{a}_1 &= \bar{K}^{-1}\left(\hat{f} - J\bar{a}_2\right),\label{eq:prev2}
    \end{align}
\end{subequations}
in which the first equation is solved by using GMRES whereas the second one is
solved using a direct solver.

Since $J$ is a submatrix of $\hat{K}$, it can be expressed in terms of $\hat{K}$ as
\begin{equation}
	J = \hat{K}\bar{H}.\label{eq:J}
\end{equation}
Substituting Equation~\ref{eq:update} into Equation~\ref{eq:J} yields
\begin{align}
	J &= \left(\bar{K}-\bar{H} \bar{E}\bar{H}^\top\right)\bar{H}\nonumber\\
	  &= \bar{K}\bar{H} - \bar{H}\bar{E}.\label{eq:KHHE}
\end{align}
Substituting Equation~\ref{eq:KHHE} into Equations~\ref{eq:prev1} and~\ref{eq:prev2}
yields
\begin{subequations}
	\begin{align}
		\left(I-\bar{H}^\top\bar{K}^{-1}\bar{H}\bar{E}\right)\bar{a}_2
		    &= \bar{H}^\top\bar{K}^{-1}\hat{f}\quad\text{and}\\
		\bar{a}_1 &= \bar{K}^{-1}\hat{f}
		    - \bar{H}\bar{a}_2+\bar{K}^{-1}\bar{H}\bar{E}\bar{a}_2,\label{eq:prev2sub}
    \end{align}
\end{subequations}
in which the first equation is identical to Equation~\ref{eq:s2}. Substituting
Equation~\ref{eq:s2b1} into Equation~\ref{eq:prev2sub} yields
\begin{equation}
	\bar{a}_1 = \bar{K}^{-1}\hat{f}
		- \bar{H}\bar{a}_2+\bar{K}^{-1}\bar{H}
		\left[\bar{H}^\top\left(\bar{f}-\hat{f}\right)-\bar{a}_3\right],
\end{equation}
which is identical to Equation~\ref{eq:augb1} if the condition in Equation~\ref{eq:fcond}
is satisfied. Hence, the two augmented formulations are mathematically equivalent.

\section{Results}
\label{sec:results}
The augmented matrix method for principal submatrix updates was evaluated
through finite element cutting experiments with five model types. This 
section provides relevant implementation details and presents experimental 
data. We compare the performances of the following three approaches:
\begin{itemize}
\item AMPS algorithm presented in Section~\ref{sec:methods}; 
\item Unsymmetric augmented matrix methods presented in \cite{yeung16} using a GMRES
    iterative solver, without preconditioning, and with two kinds of preconditioners:
    sparse approximate inverse (SPAI) and the diagonal matrix $D$ from the initial
    $LDL^{\top}$ factorization of the initial stiffness matrix; and
\item Jacobi preconditioned or nonpreconditioned conjugate gradient (CG)
    iterative solver applied on $\widehat{A}\widehat{x}=\widehat{b}$.
\end{itemize}
For the latter two approaches, only the best performing versions are included
in the figures.

\subsection{Implementation}
All experiments were conducted on a compute node with two 16-core Intel 
Xeon Processors E5-2698 v3 (``Haswell'') at 2.3 GHz, and each core equipped
with 64 KB L1 cache (32 KB instruction cache, 32 KB data cache) and 256 KB
L2 cache; as well as a 40-MB shared L3 cache per socket. In addition, there
are 128 GB DDR4 2133 MHz memory. All data represent times  averaged over
20 runs  unless overall time exceeds 30 minutes, in which case we averaged over 10 runs.

The precomputed LDL$^\top$ factorizations of the stiffness matrices were
computed using OBLIO, a sparse direct solver library \cite{dobrian}. All 
other basic linear algebra subroutines including matrix-vector products, 
dense matrix factorization and solves, as well as the GMRES iterative solver 
used in the unsymmetric augmented matrix methods and the CG solver used for 
comparison purposes were from the Intel Math Kernel Library (MKL) 
\cite{mkl}. The remainder of the code, including the computation of the
closure in $K$ induced by $H_{\delta t}$, the matrices $V_{\Delta t}$ and
$\tril{H_{\Delta t} K^{-1} H_t}$ in Equation~\ref{eq:KtHp}, and the overall
algorithm, was written by the authors.

Since the closure of a set of indices in the graph of a triangular
matrix can be found effectively column by column, and OBLIO uses supernodes 
in matrix factorization, the matrices $K$, $L$ and $V$ were stored in 
compressed sparse column matrix (CSC) format for efficient column access.
The diagonal matrix $D$ is stored in a vector of size $n$. The principal 
submatrix update $E$, the principal submatrix of the inverse
$H^\top K^{-1} H$ and the Schur complement $S_2$ were stored in dense matrix 
format for fast computations. The matrix $H$ and its transpose were
represented as an array of indices and their multiplications with other 
matrices were done by index mappings. All vectors were stored in dense 
format.

\subsection{Model Meshes}
\label{Sec:Meshes}
\begin{figure}
	\centering
	\includegraphics[scale=0.3]{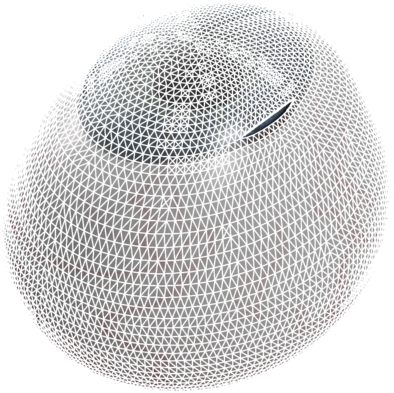}
    \caption{Rendering of the eye mesh}
    \label{fig:mesh}
\end{figure}

\begin{table*}[!htbp]
	\begin{tabularx}{\textwidth}{*{3}c{>{\centering\arraybackslash}X}}
		\toprule
		Mesh & $|V|$ & Estimated condition number & Factorization time (s) \\\midrule
		Beam & $100 - 25,600$ & $1.14 \times 10^{3} - 3.29 \times 10^{12}$ &
		    $0.02 - 1.62$\\
		Brick & $250 - 18,081$ & $2.19 \times 10^{3} - 1.18 \times 10^{5}$ &
		    $0.1 - 5.42$\\
		Eye & $17,821$ & $7.73 \times 10^6$ & $1.6$\\
		Brain & $50,737$ & failed to estimate & $7.77$\\
		\bottomrule
	\end{tabularx}
	\caption{Numerical properties of the meshes.}
	\label{tab:properties}
\end{table*}

\begin{figure}[!htbp]
    \begin{tikzpicture}
		\begin{semilogyaxis}[
	       two plot,
	       xlabel = {index (sorted)},
	       ylabel = {eigenvalue}
		]
			\addplot[midnight,mark=o] table [x=index,y=eigs] {figs/eye_small_eigs.txt};
		\end{semilogyaxis}
	\end{tikzpicture}
	\caption{Eigenspectrum of the eye mesh of $4,444$ nodes.}
	\label{fig:eye_small_eigs}
\end{figure}
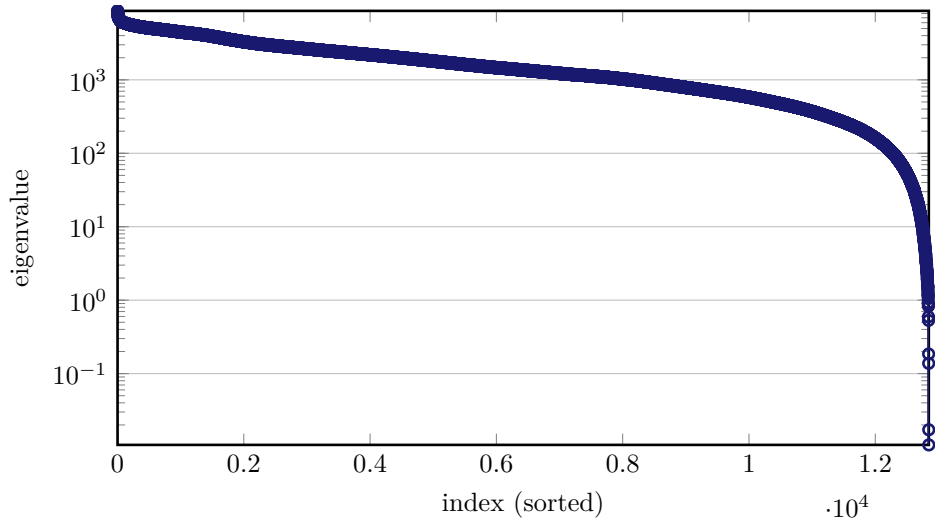

Four types of solid tetrahedral meshes were used for performance evaluation. The eye
mesh rendering is shown in Figures~\ref{fig:mesh}~and the renderings of other meshes
can be found in~\cite{yeung16}. Table~\ref{tab:properties} lists for each mesh its
number of vertices, the estimated condition number computed using Matlab's
\texttt{condest} function, and the factorization times computed using OBLIO. Since the
models are 3-dimensional, the total degrees of freedom (DOFs) in each system are 3 times
the number of vertices minus the DOFs constrained by the Dirichlet boundary conditions
$\mathcal{D}$, which is also the dimension of the matrix, i.e.,
$n = 3|V| - |\mathcal{D}|$.

\begin{enumerate}
\item \emph{Elongated Beam:}
	A group of five elongated rectangular solids with varying lengths were generated.
	Nodes were placed at regularly spaced grid points on a $5 \times 5 \times h$ grid,
	where  $h$ ranged from 4 to 1024. Each block mesh was anchored at one end of the
	solid. All elements had good aspect ratios and were arranged in a regular pattern.
	However models with greater degrees of elongation produced more poorly conditioned 
	systems of equations, as fixation at only one end meant that longer structures were 
	less stable. Thus experiments with this group of meshes illuminates the way solver 
	performance varies with stiffness matrix conditioning.
\item\emph{Brick:}
	A group of five rectangular brick solids with varying mesh resolutions were
	generated. Each of the models had the same compact physical dimension of
	$1 \times 1 \times 2$. An initial good-quality mesh was uniformly subdivided to
	produce meshes of increasingly fine resolution. These meshes allowed us to examine
	solver performance relative to node count for fixed model geometry. Similar to the
	beam meshes, zero-displacement boundary conditions were applied to one face of the
	block.
\item \emph{Eye:}
    A human eye model \cite{crouch_eye} with a clear corneal cataract incision was used
    in a simulation of corrective surgery for astigmatism. Zero displacement boundary
    conditions were applied to the posterior portion of the globe. 
    Figure~\ref{fig:eye_small_eigs} shows the eigenspectrum of an eye mesh of $4,444$
    nodes, a downsampled mesh of the eye model.
\item \emph{Brain:}
    A human brain model (contributed by INRIA to the AIM@SHAPE Shape Repository) was
    used to demonstrate applicability to surgical simulation on an organ of 
    complicated structure. Zero displacement boundary conditions were applied to the
    interior portion of the brain. The condition number could not be estimated with
    Matlab due to insufficient memory.
\end{enumerate}

On average, the nodes in the brick meshes have a higher degree of connectivity than
those in the elongated beam meshes.  This is due to a greater proportion of surface
nodes present in the beam models versus interior nodes in the brick models. The
increased connectivity leads to a higher percentage of nonzeros in the stiffness
matrix factors and larger sizes for the closures referenced in Table~\ref{tab:summary}.
These differences have a significant impact on the relative performance of the solution
methods.

\subsection{Experiments}
Performance was examined through two types of experiments: deformation of intact
meshes through changes in boundary conditions, and deformation of meshes undergoing
cutting.

\subsubsection{Deformation of Intact Meshes}
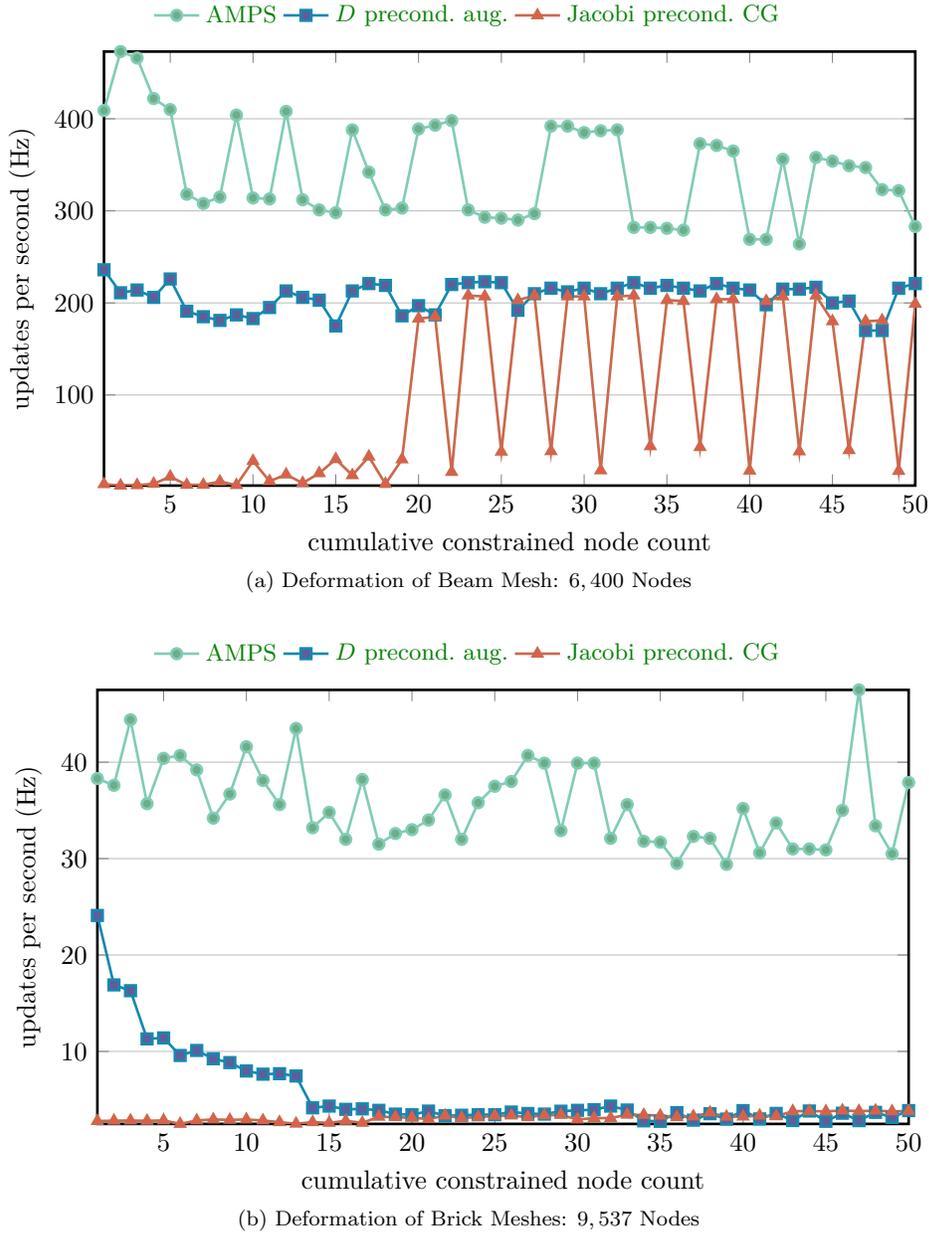
\begin{figure*}[!htbp]
    \centering
    \subfloat[Deformation of Beam Mesh: $6,400$ Nodes]{\label{fig:beam_deform}
		\begin{tikzpicture}
			\begin{axis}[
				deform timing plot,
		    	legend to name=beamdeformlegend,
		    ]
			\addplot table [x=node,y=amps_t] {figs/beam_6400_deform.txt};
			\addplot table [x=node,y=paug_t] {figs/beam_6400_deform.txt};
			\addplot table [x=node,y=pcg_t] {figs/beam_6400_deform.txt};
			\end{axis}
			\node[anchor=south] at (current bounding box.north) {\ref{beamdeformlegend}};
		\end{tikzpicture}
    }\\
    \subfloat[Deformation of Brick Meshes: $9,537$ Nodes]{\label{fig:brick_deform}
		\begin{tikzpicture}
			\begin{axis}[
		    	deform timing plot,
				legend to name=brickdeformlegend,
			]
			\addplot table [x=node,y=amps_t] {figs/brick_9537_deform.txt};
			\addplot table [x=node,y=paug_t] {figs/brick_9537_deform.txt};
			\addplot table [x=node,y=pcg_t] {figs/brick_9537_deform.txt};
			\end{axis}
			\node[anchor=south] at (current bounding box.north) {\ref{brickdeformlegend}};
		\end{tikzpicture}
	}
    \caption{Deformation update rates are shown for AMPS, the preconditioned
    augmented and CG methods as constraints are progressively added to an
    increasing number of nodes in \protect\subref{fig:beam_deform} beam and
    \protect\subref{fig:brick_deform} brick meshes.}
    \label{fig:beam_brick_deform}
\end{figure*}

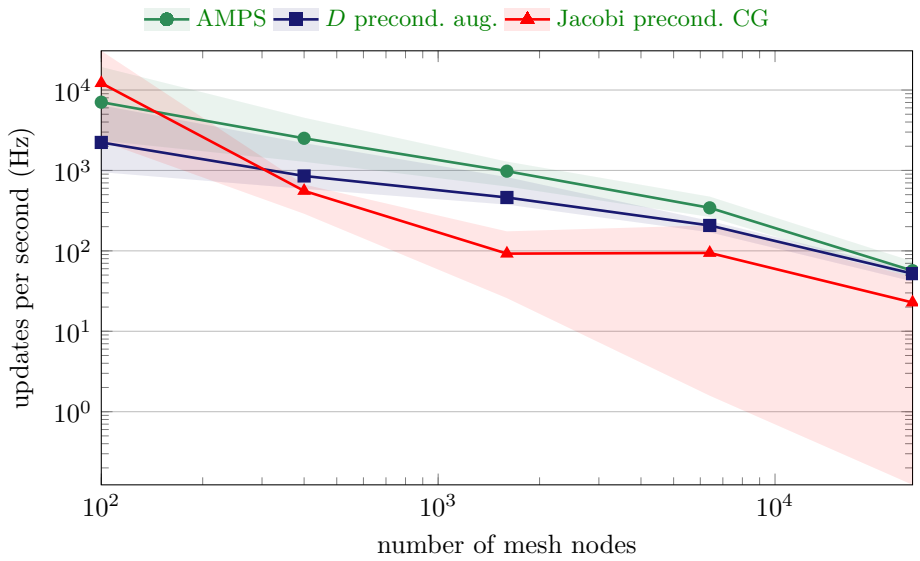
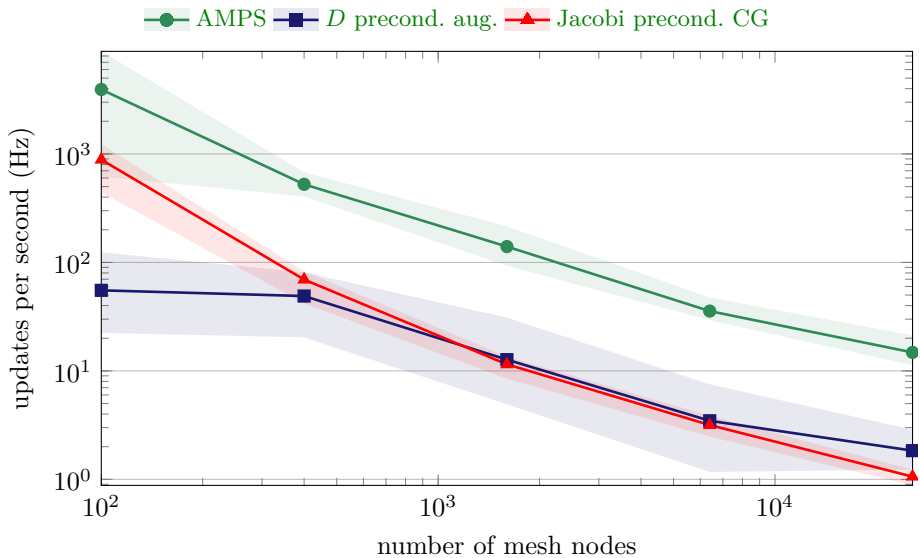
\begin{figure*}[!htbp]
	\centering
    \subfloat[Deformation of Beam Meshes]{\label{fig:beam_scale_deform}
		\begin{tikzpicture}
			\begin{loglogaxis}[
				scale plot,
		    	legend to name=beamdeformscalelegend,
		    ]
				\errorband[seagreen,
				    every mark/.append style={solid,fill=seagreen},
					mark=*]
					{figs/beam_scale_deform.txt}{size}{amps_min}{amps_max}{amps_avg};
				\errorband[midnight,
					every mark/.append style={solid,fill=midnight},
					mark=square*]
					{figs/beam_scale_deform.txt}{size}{aug_min}{aug_max}{aug_avg};
				\errorband[red,
					every mark/.append style={solid,fill=red},
					mark=triangle*]
					{figs/beam_scale_deform.txt}{size}{cg_min}{cg_max}{cg_avg};
			\end{loglogaxis}
			\node[anchor=south] at (current bounding box.north)
			    {\ref{beamdeformscalelegend}};
		\end{tikzpicture}
	}\\
	\subfloat[Deformation of Brick Meshes]{\label{fig:brick_scale_deform}
		\begin{tikzpicture}
			\begin{loglogaxis}[
				scale plot,
				legend to name=brickdeformscalelegend,
			]
				\errorband[seagreen,
					every mark/.append style={solid,fill=seagreen},
					mark=*]
					{figs/brick_scale_deform.txt}{size}{amps_min}{amps_max}{amps_avg};
				\errorband[midnight,
					every mark/.append style={solid,fill=midnight},
					mark=square*]
					{figs/brick_scale_deform.txt}{size}{aug_min}{aug_max}{aug_avg};
				\errorband[red,
					every mark/.append style={solid,fill=red},
					mark=triangle*]
					{figs/brick_scale_deform.txt}{size}{cg_min}{cg_max}{cg_avg};
			\end{loglogaxis}
			\node[anchor=south] at (current bounding box.north)
			    {\ref{brickdeformscalelegend}};
		\end{tikzpicture}
	}
	\caption{Average update rates and ranges are shown for the deformation
	experiments of the series of \protect\subref{fig:beam_scale_deform} beam and
	\protect\subref{fig:brick_scale_deform} brick meshes. AMPS results are shown
	in green, SPAI preconditioned augmented method results are shown in blue, and
	Jacobi preconditinoed CG results are shown in red.}
	\label{fig:scalability_deform}
\end{figure*}
In this group of experiments, we applied an increasing number of non-zero
essential boundary conditions to mesh nodes to create deformation.
Figure.~\ref{fig:beam_brick_deform} shows how solution time varied with
the number of constrained nodes for instances of the beam and brick meshes.

For the beam mesh, AMPS outperformed the unsymmetric augmented matrix method
by a factor of 1.65 and the CG method by 3.63, while maintaining a high average
update rate of 343 Hz (updates/sec) throughout. The unsymmetric augmented matrix method
came  second, maintaining update rates around 200 Hz. CG performed
the worst, providing updates in the range of 1.6--33 Hz for the first 19 cutting
steps, and experienced a zig-zag pattern afterwards caused by the connectivity
pattern of nodes in the tetrahedral brick mesh as explained in \cite{yeung16}.
This pattern also appeared in the results of the cutting experiments of the
beam and brick meshes, as well as the eye mesh as they have a structural
pattern in the ellipsoidal shapes.

For brick meshes, AMPS vastly outperformed the unsymmetric augmented matrix
method by a factor of 6.37, and the CG method by 11.2. AMPS maintained
relatively stable average update rates at 35.6 Hz. The unsymmetric augmented matrix
method outperformed CG as constraints were applied to the first dozen nodes,
but performance degrades as the number of constrained nodes increased,
eventually resulting in similar update rates between the augmented method
and CG. Overall, the unsymmetric augmented matrix method achieved an average
update rate of 5.59 Hz while the preconditioned CG method only had an average
update rate of 3.17 Hz.

Figure~\ref{fig:scalability_deform} is a log-log plot that shows how solution
times varied for different sizes of beam and brick meshes. The lines show the
trend of the average times for various methods and the shaded areas are the
ranges of the solution times. These graphs show that AMPS ran faster than
both the augmented and CG methods for the beam meshes except for the
very smallest instance that had only 100 nodes. It can also be observed that
CG has the largest ranges among all methods especially for the larger beam
meshes. This means that the CG solution times increased a lot while the
deformation progressed. For the brick meshes, AMPS also outperformed both
the augmented and CG methods with smaller solution time ranges than the
other methods.

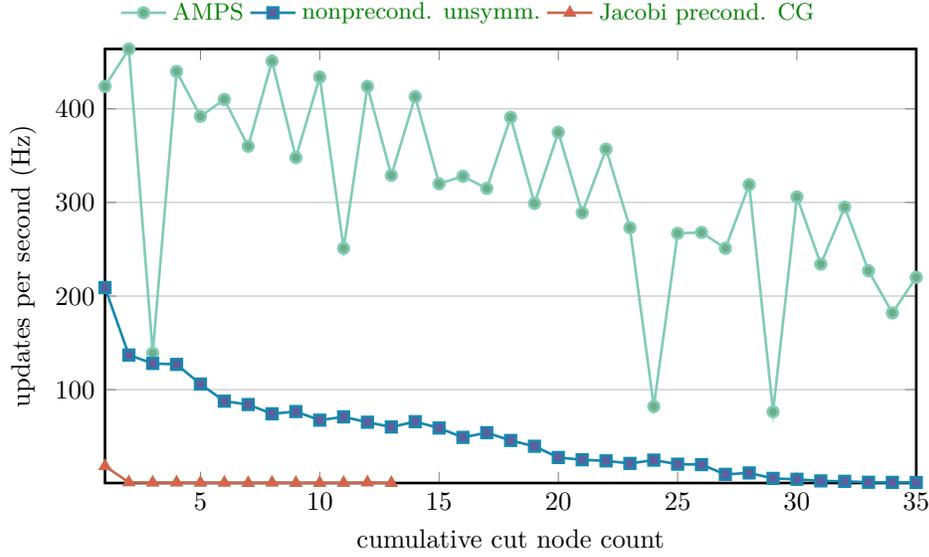
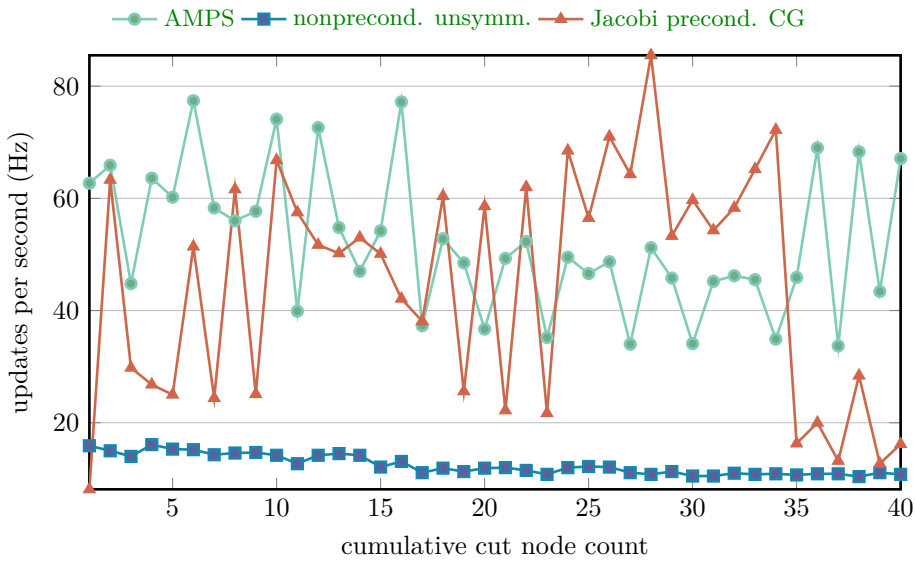
\begin{figure*}[!htbp]
    \centering
    \subfloat[Cutting of Beam Mesh: $6,400$ Nodes]{\label{fig:beam_cutting}
		\begin{tikzpicture}
			\begin{axis}[
				cutting timing plot,
				legend entries={AMPS, nonprecond. unsymm., Jacobi precond. CG},
				legend to name=beamcuttinglegend,
			]
            \pgfplotstableread{figs/beam_6400.txt}{\tabledata}
            \foreach \y in {amps_t, aug_t, pcg_t}
				\addplot+ table [x=node,y=\y] {\tabledata};
			\end{axis}
			\node at (current bounding box.north) [anchor=south] 
				{\ref{beamcuttinglegend}};
		\end{tikzpicture}
	}\\
	\subfloat[Cutting of Brick Meshes: $9,537$ Nodes]{\label{fig:brick_cutting}
		\begin{tikzpicture}
			\begin{axis}[
				cutting timing plot,
				legend entries={AMPS, nonprecond. unsymm., Jacobi precond. CG},
				legend to name=brickcuttinglegend,
			]
            \pgfplotstableread{figs/brick_9537.txt}{\tabledata}
            \foreach \y in {amps_t, aug_t, pcg_t}
				\addplot+ table [x=node,y=\y] {\tabledata};
			\end{axis}
			\node[anchor=south] at (current bounding box.north)
				{\ref{brickcuttinglegend}};
		\end{tikzpicture}
	}
  \caption{Update rates are shown for AMPS, the preconditioned augmented and CG
  methods as a cut is advanced through \protect\subref{fig:beam_cutting} a beam
  mesh and \protect\subref{fig:brick_cutting} a brick mesh.}
  \label{fig:beam_brick_cutting}
\end{figure*}

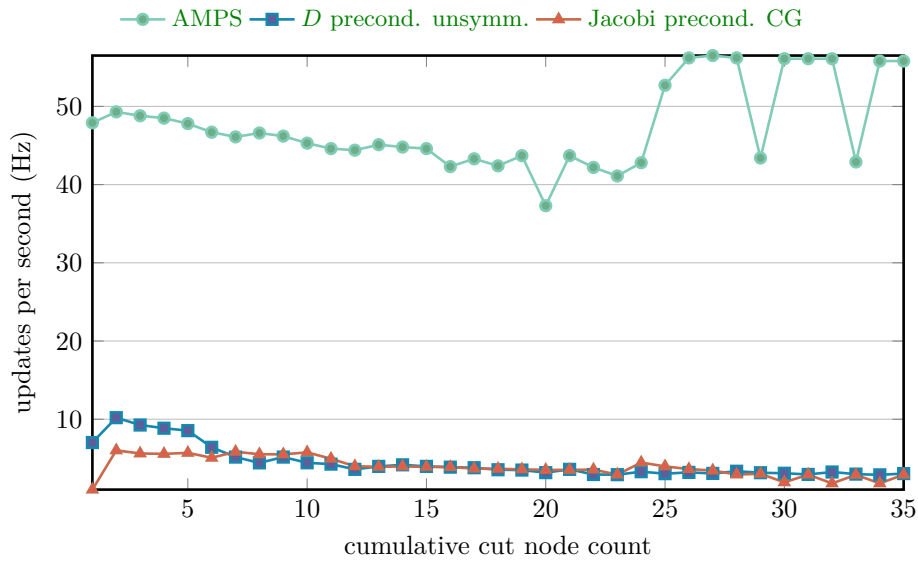
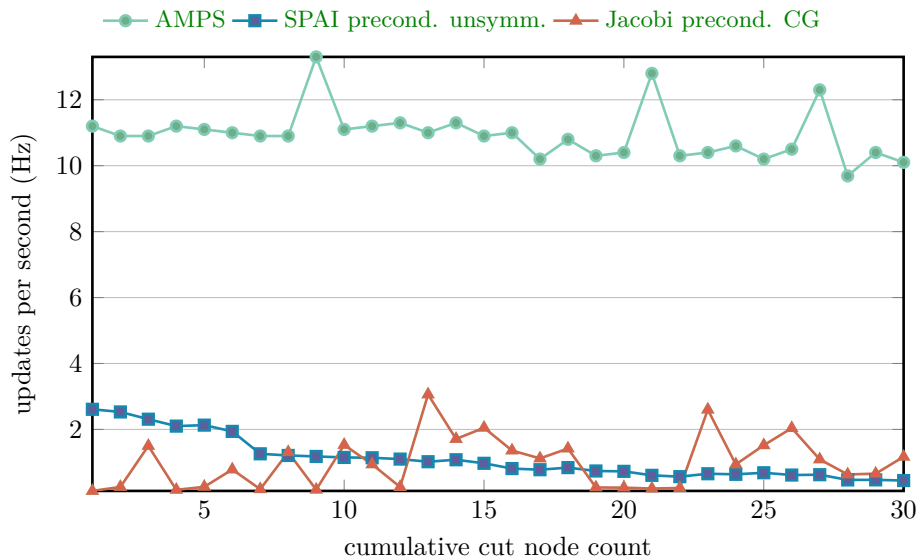
\begin{figure*}[!htbp]
    \centering
    \subfloat[Astigmatism Surgical Simulation of Eye Mesh: $17,821$ Nodes]
    {\label{fig:eye_cutting}
		\begin{tikzpicture}
			\begin{axis}[
				cutting timing plot,
		    	legend entries={AMPS, $D$ precond. unsymm., Jacobi precond. CG},
		    	legend to name=eyecuttinglegend,
		    ]
            \pgfplotstableread{figs/eye_large.txt}{\tabledata}
            \foreach \y in {amps_t, paug_t, pcg_t}
				\addplot+ table [x=node,y=\y] {\tabledata};
			\end{axis}
			\node[anchor=south] at (current bounding box.north) {\ref{eyecuttinglegend}};
		\end{tikzpicture}
	}\\
	\subfloat[Cutting of Brain Meshes: $50,737$ Nodes]
	{\label{fig:brain_cutting}
		\begin{tikzpicture}
			\begin{axis}[
				cutting timing plot,
		    	legend entries={AMPS, SPAI precond. unsymm., Jacobi precond. CG},
		    	legend to name=braincuttinglegend,
		    ]
            \pgfplotstableread{figs/brain_large.txt}{\tabledata}
            \foreach \y in {amps_t, spai_t, pcg_t}
				\addplot+ table [x=node,y=\y] {\tabledata};
			\end{axis}
			\node[anchor=south] at (current bounding box.north) {\ref{braincuttinglegend}};
		\end{tikzpicture}
    }
  \caption{Timing results are provided for \protect\subref{fig:eye_cutting} the
  eye mesh of $17,821$ nodes and \protect\subref{fig:brain_cutting} the brain mesh
  of $50,737$ nodes.}
  \label{fig:eye_brain}
\end{figure*}

\begin{figure*}[!htbp]
	\begin{tikzpicture}
		\begin{axis}[
			width=0.95*\textwidth,
			height=0.35*\textheight,
			ymajorgrids,
			align=center,
			stack plots=y,
			enlargelimits=false,
			clip=false,
			area style,
			cycle list name=breakdowncolorlist,
			yticklabel={\pgfmathparse{\tick*100}\pgfmathprintnumber{\pgfmathresult}\%},
			legend to name=bunnybreakdownlegend,
			legend columns=3,
			legend cell align={left},
		    legend style={font=\small,draw=none, at={(current bounding box.north)}, anchor=south},
			legend entries={Compute $\tril{H_{\Delta t}^\top K^{-1} H_t}$, Compute $S_2$, Form $RHS$, Solve for $\mathring{a}_2$, Solve for $\hat{a}$},
			]
			\pgfplotstableread{figs/brain_large_distribution.txt}{\tabledata}
            \foreach \y in {W, M, RHS, z, x}
			    \addplot+ table [x=cut,y=\y] {\tabledata} \closedcycle;
			\end{axis}
			\node[anchor=south] at (current bounding box.north)
			    {\ref{bunnybreakdownlegend}};
		\end{tikzpicture}
    \caption{The breakdown of computation time to steps of AMPS for the brain mesh of 50,737 nodes.}
	\label{fig:breakdown}
\end{figure*}
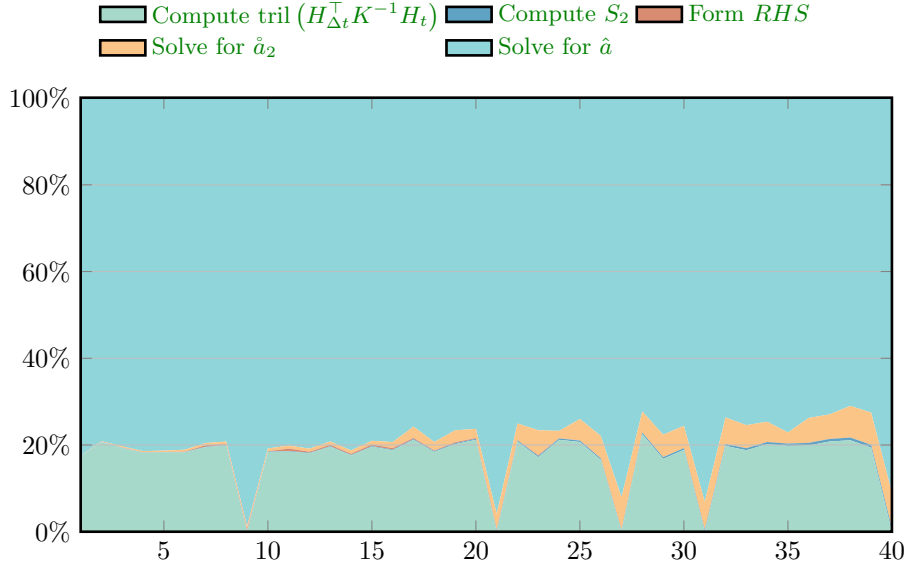
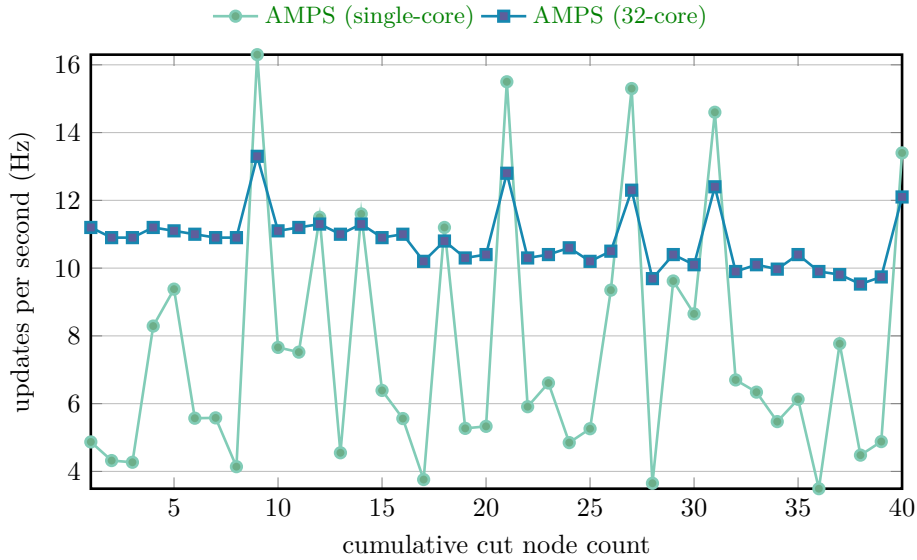
\begin{figure*}
    \begin{tikzpicture}
    	\begin{axis}[
	    	cutting speedup plot,
            legend to name=brainspeeduplegend,
	    ]
        \pgfplotstableread{figs/brain_large_speedup.txt}{\tabledata}
        \foreach \y in {single_t, multicore_t}
	        \addplot+ table [x=node,y=\y] {\tabledata};
        \end{axis}
        \node[anchor=south] at (current bounding box.north)
		    {\ref{brainspeeduplegend}};
    \end{tikzpicture}
    \caption{Single-core and 32-core results are provided for the brain mesh of 50,737 nodes.}
    \label{fig:brain_speedup}
\end{figure*}

\subsubsection{Deformation of Meshes Undergoing Cutting}
In this group of experiments we made an advancing planar cut into the volume of each
mesh. As a cut progressed, a copy of each node along the cut path was added to the mesh,
and connectivity was modified so that elements on opposite sides of the cut became
separated. The newly added node causes the linear system to increase in dimension, and
the remeshing associated with the duplicated node and all its neighboring nodes results
in a principal submatrix update to the stiffness matrix. In the results, the cut node 
count corresponds to the number of duplicated nodes resulting from the cut. Opposing
force vectors were applied to selected surface nodes to pull the cut faces apart.
Figure~\ref{fig:mesh} shows the the eye mesh at the initial stages of cutting.

While the other methods behaved differently for the cutting and deformation experiments
for the beam and brick meshes, AMPS performed similarly between the two experiments as
shown in Figure~\ref{fig:beam_cutting} compared to Figure~\ref{fig:beam_brick_deform}.
AMPS outperformed the nonpreconditioned unsymmetric augmented matrix method by a factor
of 6.06, and Jacobi preconditioned CG method by 216 in the beam cutting experiments,
providing updates in the range 167--479 Hz. The unsymmetric augmented method provided
0.83--209 Hz whereas preconditioned CG needed more than 1 second for most of the cutting
steps except for the first one, and failed to converge to any solution after the 18th
step. $D$ preconditioned and SPAI preconditioned variants ran $15.6$ and $12.5$ times
slower than AMPS respectively. On the other hand, AMPS performed on par with CG for the
brick mesh cutting experiment, providing 52.2 Hz and 44.8 Hz update rates; while the
unsymmetric augmented matrix method underperformed for this mesh, providing only an
average of 12.5 Hz update rate, as shown in Figure~\ref{fig:brick_cutting}. The $D$
preconditioned and SPAI preconditioned variants ran $11.4$ and $13.1$ times slower than
AMPS for the cutting of the brick mesh.

%
Figure~\ref{fig:breakdown} shows the breakdown of the solution times for
individual steps of the AMPS algorithm for the brain mesh of $50,737$ nodes.
The most computational expensive step was the triangular solve for the final
solution $\hat{a}$, accounting for roughly $80\%$ of the time, followed by the
computation of the principal submatrix of the inverse, accounting for roughly
$20\%$ of the time. The remaining steps are less significant. The valleys in the
area plot are due to the fact that at some cuts no additional neighboring vertices
were included in $\mathbb{H}_{\Delta t}$ and thus $\tril{H_{\Delta t}^\top K^{-1} H_t}$
is empty and the principal submatrix of the inverse of $K$ need not be updated.

Results from the eye and brain mesh cutting experiments are shown in 
Figure~\ref{fig:eye_brain}. Here we show that for the astigmatism surgical
simulation experiment AMPS vastly outperformed the $D$ preconditioned unsymmetric
augmented matrix method by a factor of $10.8$ and Jacobi preconditioned CG method by
$12.2$. For the brain model, AMPS ran $10.2$ times faster than the SPAI preconditioned 
unsymmetric augmented matrix method, $18.5$ times faster than the $D$ preconditioned
variant, $11.5$ times faster than the nonpreconditioned variant, and $11.5$ times faster
than Jacobi preconditioned CG method. The average update rates of $47.5$ Hz and
$11.4$ Hz achieved by AMPS on both the eye and brain meshes respectively make
interactive stimulation feasible.

Figure~\ref{fig:brain_speedup} shows the brain mesh cutting experiments using
AMPS on a single core versus 32 cores. Speedups vary for different cuts due to the
various numbers of new neighboring nodes of the node being cut. For cuts that
do not involve new neighboring nodes, the single-core results are even better
than those using 32 cores due to the multi-core overheads. The geometric mean
of the speedups is $1.58$.

\begin{table*}[!ht]
	\begin{tabularx}{\textwidth}{*{2}c*{3}{>{\centering\arraybackslash}X}}
		\toprule
		Mesh & $|V|$ & AMPS & SPAI precond. unsymm. aug. & Jacobi precond. CG \\\midrule
		Beam & 25,600 & $7 \times 10^{-11}$ & $1 \times 10^{-4} (10^{-4})$ & failed to converge\\
		Brick & 18,081 & $3 \times 10^{-14}$ & $5 \times 10^{-5} (10^{-5})$ & $5 \times 10^{-5} (10^{-8})$\\
		Eye & 17,821 & $4 \times 10^{-14}$ & $7 \times 10^{-4} (10^{-5})$ & $1 \times 10^{-5} (10^{-7})$\\
		Brain & 50,737 & $9 \times 10^{-14}$ & $7 \times 10^{-5} (10^{-4})$ & $1 \times 10^{-5} (10^{-5})$\\
		\bottomrule
	\end{tabularx}
	\caption{Comparison of relative residual norms 
	    ($\|\hat{K}\hat{a}-\hat{f}\|_2/\|\hat{f}\|_2$). Absolute tolerances for the
	    iterative solvers are listed in parentheses.}
	\label{tab:residual}
\end{table*}

Since AMPS uses direct solver in both augmented part and the whole solutions,
the solution accuracy of AMPS is only affected by the rounding errors amplified
by the matrix condition number. Hence, AMPS not only provided faster update times
than both the unsymmetric augmented matrix method and CG methods, but also higher
accuracy. Table~\ref{tab:residual} compares the relative residual norms of the computed
solutions of the tested methods. The absolute tolerances listed were set such that
the computed relative residual norms were less than $10^{-3}$. If lower tolerances were
set, the number of iterations and thus the solution time would increase. It can be
observed that the solutions computed by AMPS are much more accurate than the others.

\section{Conclusions and Future Work}
\label{sec:conclusion}
When meshes are cut, new nodes and elements are inserted during 
the remeshing, and new boundary conditions are imposed. 
These changes result in  principal submatrix
updates  to the stiffness system of equations, and we have demonstrated
that the solutions of the modified systems can be computed in
real-time with high accuracy even for large meshes. Our new AMPS algorithm
has outperformed an earlier  unsymmetric augmented method and CG in almost every deformation and cutting experiment. We have also observed that unlike the unsymmetric augmented method, the update rates of AMPS do not deteriorate
while the number of constrained nodes increases, or the cutting is being advanced in the meshes. These properties of AMPS are crucial for making real-time surgical simulation feasible as it requires accurate, fast and stable updates to the meshes. Refactorization would not be needed when AMPS is applied.

As we observed from the experimental results, the computation time for the
augmentation is no longer the dominating factor of the total solution time
for large meshes. More time was spent on the triangular solves in the solution. Hence, in the future one could incorporate the parallelization of the triangular solves into the AMPS algorithm. For more complicated and larger meshes, GPU and distributed parallelism could also explored.

The surgical simulations community has found the linear elastic model 
to be useful for biomechanical modeling when deformations are small and limited 
forces are applied, although linear elasticity does not adequately model organs and tissue types under heavier loading scenarios. 
Nonlinear models are not considered in this article, but 
could be investigated in the future for a broader range of surgical simulation problems, since there is evidence that viscoelastic and hyperelastic material 
models are often appropriate for modeling soft tissues \cite{fung} 
\cite{lapeer} \cite{marchesseau}.



\bibliographystyle{siamplain}
\bibliography{2018-AMPS}
\end{document}